\documentclass{article}

\usepackage{amsmath}
\usepackage{amstext}
\usepackage{amssymb}
\usepackage{color}
\usepackage{algorithmic,algorithm}
\usepackage{bm}
\usepackage{xspace}
\usepackage{color-edits}
\usepackage{fullpage}
\usepackage{graphicx}

\usepackage[font=bf,skip=3pt]{caption}

\addauthor{xh}{blue}
\addauthor{dk}{red}

\newcommand{\InfMax}{{Influence Maximization}\xspace}
\newcommand{\InfDiffMax}{{Influence Difference Maximization}\xspace}
\newcommand{\ICModel}{{Independent Cascade Model}\xspace}
\newcommand{\LTModel}{{Linear Threshold Model}\xspace}
\newcommand{\GTModel}{{Generalized Threshold Model}\xspace}

\newcommand{\Process}[3]{\bm{P}^{\text{#1}}_{#2}(#3)}
\newcommand{\ActSD}[2]{#1_{#2}}

\newcommand{\ActProb}{p}
\newcommand{\ActProbD}[1]{\ActProb_{#1}}

\newcommand{\ActProbP}{p'}
\newcommand{\ActProbDP}[1]{\ActProbP_{#1}}

\newcommand{\EdgeW}{c}
\newcommand{\EdgeWD}[1]{\EdgeW_{#1}}

\newcommand{\ParamValue}[1]{\theta_{#1}}

\newcommand{\AllParamValues}{\bm{\theta}}
\newcommand{\AllParamValuesP}{\bm{\theta'}}
\newcommand{\AllParamValuesObs}{\bm{\theta}}

\newcommand{\AllParamValuesMax}{\bm{\theta^+}}
\newcommand{\AllParamValuesMin}{\bm{\theta^-}}

\newcommand{\AllParamRange}{\bm{\Theta}}

\newcommand{\IMFuncSym}[1][]{\ifthenelse{\equal{#1}{}}{\sigma}{\sigma_{#1}}}
\newcommand{\IMFunc}[1]{\IMFuncSym(#1)}
\newcommand{\IMFuncD}[2]{\IMFuncSym_{#1}(#2)}
\newcommand{\IMFuncSymP}{\sigma'}
\newcommand{\IMFuncP}[1]{\IMFuncSymP(#1)}

\newcommand{\IMFuncSymMax}{\sigma^+}

\newcommand{\IMFuncSymMin}{\sigma^-}

\newcommand{\IDMFuncSym}[1][]{\ifthenelse{\equal{#1}{}}{\delta}{\delta_{#1}}}
\newcommand{\IDMFunc}[1]{\IDMFuncSym(#1)}
\newcommand{\IDMFuncD}[2]{\IDMFuncSym_{#1}(#2)}

\newcommand{\I}{I}
\newcommand{\ID}[1]{\I_{#1}}
\newcommand{\UBD}[1]{r_{#1}}
\newcommand{\LBD}[1]{\ell_{#1}}

\newcommand{\Th}{\psi}
\newcommand{\ThD}[1]{\Th_{#1}}
\newcommand{\SET}{S}
\newcommand{\SeedS}{A_0}
\newcommand{\SeedSUD}[2]{A^{#1}_{#2}}
\newcommand{\OptSD}[1]{\SeedSUD{*}{#1}}

\newcommand{\ExpInfDet}{\Delta}

\newcommand{\ExpEdgeCD}[1]{c_{#1}}
\newcommand{\Scaling}{p}

\newcommand{\Set}[1]{\{#1\}}
\newcommand{\Expt}[1]{\mathbb{E}[#1]}

\DeclareMathOperator{\argmax}{argmax}
\providecommand{\Kth}[1]{\ensuremath{{#1}^{\rm th}}}
\providecommand{\Norm}[2][]{\ensuremath{%
\ifthenelse{\equal{#1}{}}{\|{#2}\|}{\|{#2}\|_{{#1}}}}\xspace}

\def\QED{{\phantom{x}} \hfill \ensuremath{\rule{1.3ex}{1.3ex}}}

\providecommand{\Expect}[2][]{\ensuremath{%
\ifthenelse{\equal{#1}{}}{\mathbb{E}}{\mathbb{E}_{#1}}%
\left[#2\right]}\xspace}

\providecommand{\PROB}{\ensuremath{{\rm Prob}}\xspace}
\providecommand{\Prob}[2][]{\ensuremath{%
\ifthenelse{\equal{#1}{}}{\PROB[#2]}{\PROB_{#1}[#2]}}\xspace}

\newtheorem{theorem}{Theorem}

\newtheorem{definition}{Definition}

\begin{document}

\title{Stability of Influence Maximization}

\author{Xinran He
\thanks{Department of Computer Science, University of Southern California;
xinranhe@usc.edu}
\and
David Kempe
\thanks{Department of Computer Science, University of Southern California;
dkempe@usc.edu}
}

\maketitle
\begin{abstract}
The present article serves as an erratum to our paper of the same
title, which was presented and published in the KDD 2014 conference.
In that article, we claimed falsely that the objective function
defined in Section~\ref{sec:intro:infdiffmax} is non-monotone
submodular. We are deeply indebted to Debmalya Mandal, Jean
Pouget-Abadie and Yaron Singer for bringing to our attention a
counter-example to that claim.

Subsequent to becoming aware of the counter-example, we have shown
that the objective function is in fact NP-hard to approximate to
within a factor of $O(n^{1-\epsilon})$ for any $\epsilon > 0$.

In an attempt to fix the record, the present article combines the
problem motivation, models, and experimental results sections from the
original incorrect article with the new hardness result.
We would like readers to only cite and use this version (which will
remain an unpublished note) instead of the incorrect conference version.
\end{abstract}



\newpage

\section{Introduction}
The processes and dynamics by which information and behaviors spread
through social networks have long interested scientists within many
areas.
Understanding such processes has the potential to shed light on human
social structure, and to impact the strategies used to promote
behaviors or products.
While the interest in the subject is long-standing, recent
increased availability of social network and information diffusion
data (through sites such as Facebook, Twitter, and LinkedIn) has
raised the prospect of applying social network analysis at a large
scale to positive effect.
Consequently, the resulting algorithmic questions have received
widespread interest in the computer science community.

Among the broad algorithmic domains, \InfMax has been repeatedly held
up as having the potential to be of societal and financial value.
The high-level hope is that based on observed data --- such as social
network information and past behavior --- an
algorithm could infer which individuals are likely to influence which
others. This information could in turn be used to effect desired behavior,
such as refraining from smoking, using superior crops, or purchasing a
product.
In the latter case, the goal of effecting desired behavior is usually
termed \emph{viral marketing}.

Consequently, both the problem of inferring the influence between
individuals
\cite{gomez-rodriguez:balduzzi:schoelkopf:uncovering,gomez-rodriguez:leskovec:krause:inferring,gomez-rodriguez:schoelkopf:multiple,goyal:bonchi:lakshmanan:learning,myers:leskovec:convexity}
and that of maximizing the spread of a desired behavior
have been studied extensively.
For the \InfMax problem, a large number of models have been proposed,
along with many heuristics with and without approximation guarantees
\cite{borgs:brautbar:chayes:lucier:nearly-optimal,chen:wang:yang:efficient,chen:yuan:zhang:scalable,InfluenceSpread,InfluenceSpreadICALP,khanna:lucier:influence-maximization,mossel:roch:submodular,chen:wang:wang:prevalent,wang:cong:song:xie}.
(See the monograph \cite{chen:lakshmanan:castillo:influence-maximization-book}
for a recent overview of work in the area.)

However, one crucial aspect of the problem has --- with very few
exceptions discussed in Section~\ref{sec:adversarial-random} ---
gone largely unstudied.
Contrary to many other algorithmic domains,
\emph{noise in social network data is not an exception, but the norm}.
Indeed, one could argue that the very notion of a ``social link'' is
not properly defined in the first place, so that any representation of
a social network is only an approximation of reality.
This issue is much more pronounced for a goal such as \InfMax.
Here, the required data include, for every pair $(u,v)$ of
individuals, a numerical value for the strength of influence from $u$
to $v$ and vice versa.
This influence strength will naturally depend on context (e.g., what
exact product or behavior is being spread);
furthermore, it cannot be observed directly, and must therefore be
inferred from observed behavior or individuals' reports;
all of these are inherently very noisy.

When the inferred influence strength parameters differ from the actual
ground truth, even an optimal algorithm is bound to return suboptimal
solutions, for it will optimize the wrong objective function:
a solution that appears good with respect to the incorrect parameters
may be bad with respect to the actual ones.
If relatively small errors in the inferred parameters could lead to
highly suboptimal solutions, this would cast serious doubts on the
practical viability of algorithmic influence maximization.
\emph{Therefore, in the present paper, we begin an in-depth study of
  the effect of noise on the performance of \InfMax algorithms.}

\subsection{The \ICModel}
\label{sec:intro-model-overview}
We  study this question under two widely adopted models for influence
diffusion \cite{InfluenceSpread}:
the \emph{Independent Cascade (IC) Model}
and the \emph{Linear Threshold (LT) Model}.
Both of these models fit in the following framework:
The algorithm selects a \emph{seed set} $\SeedS$ of $k$ nodes, which
begin \emph{active} (having adopted the behavior).
Starting with $\SeedS$, the process proceeds in discrete time
steps: in each time step, according to a probabilistic process,
additional nodes may become active based on the influence from their
neighbors. Active nodes never become inactive, and the process
terminates when no new nodes become active in a time step.
The goal is to maximize the expected number of active nodes when
the process terminates;
this expected number is denoted by $\IMFunc{\SeedS}$.

To illustrate the questions and approaches, we describe the IC model
in this section.
(A formal description of the LT model and general definitions of all
concepts are given in Section~\ref{sec:models}.)
Under the IC model, the probabilistic process is
particularly simple and intuitive. When a node $u$ becomes active in
step $t$, it attempts to activate all currently inactive neighbors in
step $t+1$.
For each neighbor $v$, it succeeds with a known probability
$\ActProbD{u,v}$.
If it succeeds, $v$ becomes active; otherwise, $v$ remains inactive.
Once $u$ has made all these attempts, it does not get to make further
activation attempts at later times.
It was shown in \cite{InfluenceSpread}
that the set of nodes active at the end can be characterized
alternatively as follows: for each ordered pair $(u,v)$ independently,
insert the directed edge $(u,v)$ with probability $\ActProbD{u,v}$.
Then, the active nodes are exactly the ones reachable via directed
paths from $\SeedS$.

\subsection{Can Instability Occur?}
\label{sec:instability-example}
Suppose that we have inferred all parameters $\ActProbD{u,v}$, but are
concerned that they may be slightly off: in reality, the
influence probabilities are $\ActProbDP{u,v} \approx \ActProbD{u,v}$.
Are there instances in which a seed set $\SeedS$ that is very influential
with respect to the $\ActProbD{u,v}$ may be much less influential with respect
to the $\ActProbDP{u,v}$?
It is natural to suspect that this might not occur:
when the objective function $\IMFuncSym$ varies
sufficiently smoothly with the input parameters (e.g., for linear
objectives), small changes in the parameters only lead to small
changes in the objective value; therefore, optimizing with respect
to a perturbed input still leads to a near-optimal solution.

However, the objective $\IMFuncSym$ of \InfMax does not depend on the
parameters in a smooth way.
To illustrate the issues at play, consider the following instance of
the IC model.
The social network consists of two disjoint bidirected cliques $K_n$,
and $\ActProbD{u,v} = \hat{\ActProb}$ for all $u,v$ in the same clique;
in other words, for each directed edge, the same activation probability
$\hat{\ActProb}$ is observed.
The algorithm gets to select exactly $k=1$ node.
Notice that because all nodes look the same, any algorithm essentially
chooses an arbitrary node, which may as well be from Clique 1.

Let $\hat{\ActProb}=1/n$ be the sharp
threshold for the emergence of a giant component in the
Erd\H{o}s-R\'{e}nyi Random Graph $G(n,p)$.
It is well known \cite{bollobas:random-graphs,erdos:renyi:gnp} that
the largest connected component of $G(n,p)$ has size
$O(\log n)$ for any $p \leq \hat{\ActProb} - \Omega(1/n)$, and
size $\Omega(n)$ for any $p \geq \hat{\ActProb} + \Omega(1/n)$.
Thus, if unbeknownst to the algorithm, all true activation
probabilities in Clique 1 are $p \leq \hat{\ActProb} - \Omega(1/n)$,
while all true activation probabilities in Clique 2 are
$p \geq \hat{\ActProb} + \Omega(1/n)$, the algorithm only activates
$O(\log n)$ nodes in expectation, while it could have reached
$\Omega(n)$ nodes by choosing Clique 2.
Hence, small adversarial perturbations to the input
parameters can lead to highly suboptimal solutions from any algorithm.\footnote{The example reveals a close connection between the
  stability of an IC instance and the question whether a uniform activation
  probability $\ActProb$ lies close to the edge percolation threshold
  of the underlying graph.
  Characterizing the percolation threshold of families of graphs has been
  a notoriously hard problem. Successful characterizations have only
  been obtained for very few specific classes (such as $d$-dimensional grids
  \cite{kesten:asymptotics} and $d$-regular expander graphs
  \cite{alon:benjamini:stacey}). Therefore, it is unlikely
  that a clean characterization of stable and unstable instances can be
  obtained.
  The connection to percolation also reveals that the instability was not
  an artifact of having high node degrees. By the result of Alon et
  al.~\cite{alon:benjamini:stacey}, the same behavior will be obtained
  if both components are $d$-regular expander graphs, since such
  graphs also have a sharp percolation threshold.}

\subsection{Diagnosing Instability}
The example of two cliques shows that there exist
\emph{unstable instances}, in which an optimal solution to the observed
parameters is highly suboptimal when the observed parameters are
slightly perturbed compared to the true parameters.
Of course, not every instance of \InfMax is unstable:
for instance, when the probability $\hat{\ActProb}$ in the Two-Clique
instance is bounded away from the critical threshold of $G(n,p)$, the
objective function varies much more smoothly with $\hat{\ActProb}$.
This motivates the following algorithmic question, which is the main
focus of our paper:
\emph{Given an instance of \InfMax, can we diagnose efficiently
  whether it is stable or unstable?}

To make this question precise, we formulate a model of perturbations.
We assume that for each edge $(u,v)$,
in addition to the observed activation probability $\ActProbD{u,v}$,
we are given an interval $\ID{u,v} \ni \ActProbD{u,v}$
of values that the \emph{actual probability} $\ActProbDP{u,v}$ could assume.
The \emph{true values} $\ActProbDP{u,v}$ are chosen from the
intervals $\ID{u,v}$ by an adversary;
they induce an objective function $\IMFuncSymP$ which
the algorithm would like to maximize, while the observed values induce a
different objective function $\IMFuncSym$ which the algorithm
actually has access to.

An instance $(\ActProbD{u,v}, \ID{u,v})_{u,v}$ is
\emph{stable} if $|\IMFunc{\SET} - \IMFuncP{\SET}|$ is small
for all objective functions $\IMFuncSymP$ induced by legal probability
settings, and for all seed sets $\SET$ of size $k$.
Here, ``small'' is defined relative to the objective
function value $\IMFunc{\SeedS^*}$ of the optimum set.

When $|\IMFunc{\SET} - \IMFuncP{\SET}|$ is small compared to
$\IMFunc{\SeedS^*}$ for all sets $\SET$,
a user can have confidence that his optimization result will provide
decent performance guarantees even if his input was perturbed.
The converse is of course not necessarily true:
even in unstable instances, a solution that was optimal for the
observed input \emph{may} still be very good for the true input parameters.




\subsection{\InfDiffMax}
\label{sec:intro:infdiffmax}
Trying to determine whether there are a function
$\IMFuncSymP$ and a set $\SET$ for which $|\IMFunc{\SET} -
\IMFuncP{\SET}|$ is large motivates the following optimization problem:
Maximize $|\IMFunc{\SET} - \IMFuncP{\SET}|$ over all feasible
functions $\IMFuncSymP$ and all sets $\SET$.
For any given set $\SET$, the objective is maximized either by making
all probabilities (and thus $\IMFuncP{\SET}$) as small as possible,
or by making all probabilities (and thus $\IMFuncP{\SET}$) as large as possible.\footnote{This observation
  relies crucially on the fact that each $\ActProbD{u,v}$ can
  independently take on any value in $\ID{u,v}$. If the adversary
  were constrained by the total absolute deviation or sum of squares
  of deviations of parameters, this would no longer be the case.
  This issue is discussed in Section~\ref{sec:conclusions}.}
We denote the resulting two objective functions by
$\IMFuncSymMin$ and $\IMFuncSymMax$, respectively.
The following definition then captures the optimization goal.

\begin{definition}[\InfDiffMax]
\label{def:influence-difference}
Given two instances with probabilities $\ActProbD{u,v} \geq
\ActProbDP{u,v}$ for all $u,v$,
let $\IMFuncSym$ and $\IMFuncSymP$ be their
respective influence functions.
Find a set $\SET$ of size $k$
maximizing $\IDMFunc{\SET} := \IMFunc{\SET} - \IMFuncP{\SET}$.
\end{definition}
In this generality,
the \InfDiffMax problem subsumes the \InfMax
problem, by setting $\ActProbDP{u,v} \equiv 0$
(and thus also $\IMFuncSymP \equiv 0$).

While \InfDiffMax subsumes \InfMax, whose objective function is
monotone and submodular, the objective function of \InfDiffMax is in
general neither. To see non-monotonicity, notice that
$\IDMFunc{\emptyset} = \IDMFunc{V} = 0$, while
generally $\IDMFunc{\SET} > 0$ for some sets $\SET$.

The function is also not in general submodular, a fact brought to our
attention by Debmalya Mandal, Jean Pouget-Abadie and Yaron Singer, and
in contrast to the main result claimed in a prior version of the
present article. The following example shows non-submodularity for
both the IC and LT Models.

The graph has four nodes $V = \Set{u,v,x,y}$ and three edges
$(u,v), (v,x), (x,y)$.
The edges $(v,x)$ and $(x,y)$ are known to have an activation
probability of 1, while the edge $(u,v)$ has an adversarially chosen
activation probability in the interval $[0,1]$.
With $S = \Set{u}$ and $T = \Set{u,x}$, we obtain that
$\IDMFunc{S + v} - \IDMFunc{S} = |\emptyset| - |\Set{v,x,y}| = -3$,
while $\IDMFunc{T + v} - \IDMFunc{T} = |\emptyset| - |\Set{v}| = -1$,
which violates submodularity.

In fact, we establish a very strong hardness result here, in the form
of the following theorem, whose proof is given in Section~\ref{sec:ApxProof}.

\begin{theorem} \label{thm:apxhard}
Under the \ICModel, the \InfDiffMax objective
function $\IDMFunc{S}$ cannot be approximated better than
$n^{1-\epsilon}$ for any $\epsilon > 0$ unless
$\mbox{NP} \subseteq \mbox{ZPP}$.
\end{theorem}

\subsection{Experiments}
Next, we investigate how pervasive instabilities are in real data.
We evaluate frequently used synthetic models (2D grids,
random regular graphs, small-world networks, and preferential
attachment graphs) and real-world data sets (computer science theory
collaborations and retweets about the Haiti earthquake).
We focus on the \ICModel, and vary the influence strengths over a broad
range of commonly studied values.
We consider different relative
perturbation levels $\ExpInfDet$, ranging from $1\%$ to $50\%$.
The adversary can thus choose the actual activation probability to lie
in the interval
$[(1-\ExpInfDet) \ActProbD{u,v}, (1+\ExpInfDet) \ActProbD{u,v}]$.

To calculate a value for the maximum possible Influence
Difference, we use the random greedy algorithm of Buchbinder et
al.~\cite{Buchbinder:Feldman:Naor:Schwartz}. This choice of algorithm
was motivated by the false belief that the objective function is
submodular, in which case the algorithm would have provided a $1/e$
approximation. Notice, however, that the algorithm can only
\emph{underestimate} the maximum possible objective function value.
Thus, when the Random Greedy algorithm finds a set with large
influence difference, it suggests that the misestimations due to
parameter misestimates may drown out the objective value, rendering
Influence Maximization outputs very spurious.
On the other hand, when the objective value obtained by the Random
Greedy algorithm is small, no positive guarantees can be provided.

Our experiments suggest that perturbations can have significantly
different effects depending on the network structure and observed values.
As a general rule of thumb, perturbations above
$20\%$ relative to the parameter values could significantly distort
the optimum solution.
For smaller errors (10\% or less relative error), the values
  obtained by the algorithm are fairly small; however, as cautioned
  above, the actual deviations may still be large.

Since errors above $20\%$ should be considered quite common for
estimated social network parameters, our results suggest that
practitioners exercise care in evaluating the stability
of their problem instances, and treat the output of \InfMax
algorithms with a healthy dose of skepticism.

\subsection{Adversarial vs.~Random Perturbations}
\label{sec:adversarial-random}
One may question why we choose to study adversarial
instead of random perturbations. This choice is for three
reasons:
\begin{description}
\item[Theoretical:] Worst-case analysis provides
stronger guarantees, as it is not based on particular assumptions
about the distribution of noise.
\item[Practical:] Most random noise models assume independence of
noise across edges.
However, we believe that in practice, both the techniques used for
inferring model parameters as well as the data sources they are based
on may well exhibit systematic bias, i.e., the noise will not be
independent.
For instance, a particular subpopulation may systematically
underreport the extent to which they seek others' advice, or may have
fewer visible indicators (such as posts) revealing their behavior.
\item[Modeling Interest:] Perhaps most importantly, most natural
random noise models do not add anything to the IC and LT models.
As an illustration, consider the random noise models studied in
recent work by Goyal, Bonchi and
Lakshmanan~\cite{Goyal:Bonchi:Lakshmanan:dataInfMax}
and Adiga et al.~\cite{adiga:kuhlman:sensitivity}.
Goyal et al.~assume that for each edge $(u,v)$, the value of
$\ActProbD{u,v}$ is perturbed with uniformly random noise from a known
interval.
Adiga et al.~assume that each edge $(u,v)$ that was observed to be
present is actually absent with some probability $\epsilon$, while
each edge that was not observed is actually present with probability
$\epsilon$; in other words, each edge's presence is independently
flipped with probability $\epsilon$.

The standard \ICModel subsumes both models straightforwardly.
Suppose that a decision is to be made about whether $u$ activates
$v$. In the model of Goyal et al., we can first draw the actual
(perturbed) value of $\ActProbDP{u,v}$ from its known distribution;
subsequently, $u$ activates $v$ with probability $\ActProbDP{u,v}$;
in total, $u$ activates $v$ with probability $\Expect{\ActProbDP{u,v}}$.
Thus, we obtain an instance of the IC model in which all edge
probabilities $\ActProbD{u,v}$ are replaced by $\Expect{\ActProbDP{u,v}}$.
In the special case when the noise has mean 0, this expectation is
exactly equal to $\ActProbD{u,v}$, which explains why Goyal et
al.~observed the noise to not affect the outcome at all.

In the model of Adiga et al., we first determine whether the edge is
actually present; when it was observed present, this happens with
probability $1-\epsilon$; otherwise with probability $\epsilon$.
Subsequently, the activation succeeds with probability $p$.
(\cite{adiga:kuhlman:sensitivity} assumed uniform probabilities).
Thus, the model is an instance of the IC model in which the
activation probabilities on all observed edges are
$p(1-\epsilon)$, while those on unobserved edges are $p \epsilon$.
This reduction explains the theoretical results obtained by
Adiga et al.

More fundamentally, practically all ``natural'' random processes that
independently affect edges of the graph can be ``absorbed into'' the
activation probabilities themselves; as a result, random noise does
not at all play the result of actual noise.
\end{description}

\section{Models and Preliminaries}
\label{sec:models}

The social network is modeled by a directed graph $G=(V,E)$ on $n$
nodes. All parameters for non-existing edges are assumed to be 0.
We first describe models of influence diffusion, and then models of
parameter perturbation.

\subsection{Influence Diffusion Models}
Most of the models for \InfMax have been based on the \ICModel (see
Section~\ref{sec:intro-model-overview}) and
\LTModel studied in \cite{InfluenceSpread} and their generalizations.
Like the \ICModel, the \LTModel also proceeds in discrete rounds.
Each edge $(u,v)$ is equipped with a weight $\EdgeWD{u,v} \in [0,1]$,
satisfying $\sum_{u \to v} \EdgeWD{u,v} \leq 1$ for all nodes $v$.
(By $u \to v$, we denote that there is a directed edge $(u,v)$.)
Each node $v$ initially draws a threshold $\ThD{v}$ independently and
uniformly at random from $[0,1]$.
A set $\SeedS$ of nodes is activated at time 0,
and we use $\ActSD{A}{t}$ to denote the set of nodes active at time $t$.
In each discrete round $t$, each node $v$ checks if
$\sum_{u \in \ActSD{A}{t-1}, u \to v} \EdgeWD{u,v} \geq \ThD{v}$.
If so, $v$ becomes active at time $t$, and remains active subsequently.

Any instance of the \InfMax problem is characterized by its parameters.
For the LT model, the parameters are the $n^2$ edge weights
$\EdgeWD{u,v}$ for all edges $(u,v)$.
Similarly, for the IC model, the parameters are
the edge activation probabilities $\ActProbD{u,v}$ for all edges $(u,v)$.
To unify notation, we write
$\AllParamValues = (\ParamValue{u,v})_{(u,v) \in E}$ for the vector of
all parameter values, where $\ParamValue{u,v}$ could be either
$\EdgeWD{u,v}$ or $\ActProbD{u,v}$.

Both the IC and LT model define random processes that continue
until the diffusion process quiesces, i.e., no new activations occur.
Let $\tau \leq n$ be the (random) time at which this happens.
It is clear that $\tau \leq n$ always, since at least one more node
becomes active in each round.
We denote the stochastic process by
$\Process{Mod}{\AllParamValues}{\SeedS} = (\ActSD{A}{t})_{t=0}^{\tau}$,
with $\text{Mod} \in \Set{\text{IC}, \text{LT}}$ denoting the model.
The final set of active nodes is $\ActSD{A}{\tau}$.
We can now formally define the \InfMax problem:

\begin{definition}[\InfMax]
The \InfMax problem consists of maximizing
the objective $\IMFunc{\SeedS} := \Expt{|\ActSD{A}{\tau}|}$
(i.e., the expected number of active nodes in the end\footnote{%
Our results carry over unchanged if we assign each node a non-negative
value $r_v$, and the goal is to maximize
$\sum_{v \in \ActSD{A}{\tau}} r_v$.
We focus on the case of uniform values for notational convenience only.}),
subject to a cardinality constraint $|\SeedS| \leq k$.
\end{definition}

The key insight behind most prior work on algorithmic \InfMax is
that the objective function $\IMFunc{S}$ is a monotone and submodular
function of $S$.
This was proved for the IC and LT models in \cite{InfluenceSpread},
and subsequently for a generalization called \GTModel (proposed in
\cite{InfluenceSpread}) by Mossel and Roch \cite{mossel:roch:submodular}.



\subsection{Models for Perturbations}
\label{sec:perturbations}

To model adversarial input perturbations, we assume that
for each of the edges $(u,v)$, we are given an interval
$\ID{u,v} = [\LBD{u,v}, \UBD{u,v}] \subseteq [0,1]$ with $\ParamValue{u,v} \in \ID{u,v}$.
For the \LTModel, to ensure that the resulting activation functions
are always submodular, we require that
$\sum_{u \to v} \UBD{u,v} \leq 1$ for all nodes $v$.
We write $\AllParamRange = \times_{(u,v) \in E} \ID{u,v}$
for the set of all allowable parameter settings.
The adversary must guarantee that the ground truth parameter values
satisfy $\AllParamValuesP \in \AllParamRange$;
subject to this requirement, the adversary can choose the
actual parameter values arbitrarily.

Together, the parameter values $\AllParamValues$ determine an
instance of the \InfMax problem.
We will usually be explicit about indicating the dependence of the
objective function on the parameter setting.
We write $\IMFuncSym[\AllParamValues]$ for the objective function
obtained with parameter values $\AllParamValues$, and only omit the
parameters when they are clear from the context.
For a given setting of parameters, we will denote by
$\OptSD{\AllParamValues} \in \argmax_{S} \IMFuncD{\AllParamValues}{S}$
a solution maximizing the expected influence under parameter values
$\AllParamValues$.

\subsection{Influence Difference Maximization}
In order to capture to what extent adversarial changes in the
parameters can lead to misestimates of any set's influence, we are
interested in the quantity
\begin{equation}
\max_{\SET} \max_{\AllParamValuesP \in \AllParamRange}
| \IMFuncD{\AllParamValues}{\SET} - \IMFuncD{\AllParamValuesP}{\SET} |,
\label{eqn:general-IDM}
\end{equation}
where $\AllParamValuesObs$ denotes the observed parameter values.
For two parameter settings $\AllParamValues, \AllParamValuesP$ with
$\AllParamValues \geq \AllParamValuesP$ coordinate-wise, it is not difficult to
show using a simple coupling argument that
$\IMFuncD{\AllParamValues}{\SET} \geq \IMFuncD{\AllParamValuesP}{\SET}$ for
all $\SET$.
Therefore, for any fixed set $\SET$, the maximum is attained either
by making $\AllParamValuesP$ as large as possible or as small as possible.
Hence, solving the following problem is sufficient to maximize
\eqref{eqn:general-IDM}.

\begin{definition}
Given an influence model and two parameter settings
$\AllParamValues, \AllParamValuesP$ with
$\AllParamValues \geq \AllParamValuesP$ coordinate-wise, define
\begin{equation}
\IDMFuncD{\AllParamValues,\AllParamValuesP}{\SET}
\; = \; \IMFuncD{\AllParamValues}{\SET} - \IMFuncD{\AllParamValuesP}{\SET}.
\end{equation}
Given the set size $k$, the \InfDiffMax (IDM) problem is
defined as follows:
\begin{equation}
\begin{array}{cc}
  \text{Maximize} & \IDMFuncD{\AllParamValues, \AllParamValuesP}{\SET}\\
  \text{subject to} & |\SET| = k.
\end{array}
\end{equation}
\end{definition}

\section{Approximation hardness}
\label{sec:ApxProof}

In this section, we prove Theorem~\ref{thm:apxhard}.

\begin{extraproof}{Theorem~\ref{thm:apxhard}}

We establish the approximation hardness of \InfDiffMax without any
constraint on the cardinality of the seed set $\SeedS$.
From this version, the hardness of the constrained problem is inferred
easily as follows: if any better approximation could be obtained for
the constrained problem, one could simply enumerate over all possible
values of $k$ from $1$ to $n$, and retain the best solution, which
would yield the same approximation guarantee for the unconstrained
problem.

We give an approximation-preserving reduction from the
\textsc{Maximum Independent Set} problem to the \InfDiffMax problem.
It is well known that \textsc{Maximum Independent Set} cannot be
approximated better than $O(n^{1-\epsilon})$ for any $\epsilon > 0$
unless $\mbox{NP} \subseteq \mbox{ZPP}$ \cite{hastad:clique}.

Let $G=(V,E)$ be an instance of the \textsc{Maximum Independent Set}
problem, with $|V|=n$.
We construct from $G$ a \emph{directed} bipartite graph $G'$ with
vertex set $V' \cup V''$.
For each node $v_i \in V$, there are nodes $v'_i \in V'$
and $v''_i \in V''$.
The edge set is $E' \cup E''$, where
$E' = \Set{(v'_i, v''_j) \, | \, (v_i, v_j) \in E}$,
and $E'' = \Set{(v'_i, v''_i) \, | \, v_i \in V}$.
All edges of $E'$ are known to have an activation probability of 1,
while all edges of $E''$ have an activation probability from the
interval $[0,1]$.

The difference is maximized by making all probabilities as large for
one function (meaning that all edges in $E' \cup E''$ are present
deterministically), while making them as small as possible for the
other (meaning that exactly the edges in $E'$ are present).

First, let $S$ be an independent set in $G$. Consider the set
$S' = \Set{v'_i \, | \, v_i \in S}$.
Each node $v''_i$ with $v_i \in S$ is reachable from the corresponding
$v'_i$ in $G'$, but not in $(V' \cup V'', E')$, because $S$ is
independent. Hence, the objective function value obtained in
\InfDiffMax is at least $|S|$.

Conversely, consider an optimal solution $S'$ to the \InfDiffMax
problem.
Without loss of generality, we may assume that $S' \subseteq V'$:
any node $v''_j \in V''$ can be removed from $S'$ without lowering the
objective value.
Assume that $S := \Set{v_i \in V \, | \, v'_i \in S'}$ is not independent,
and that $(v_i, v_j) \in E$ for $v_i, v_j \in S$.
Then, removing $v'_j$ from $S'$ cannot lower the \InfDiffMax objective
value of $S'$: all of $v'_j$'s neighbors in $V''$ contribute 0, as
they are reachable using $E'$ already; furthermore, $v''_j$ also does
not contribute, as it is reachable using $E'$ from $v'_i$.
Thus, any node with a neighbor in $S$ can be removed from $S'$,
meaning that $S$ is without loss of generality independent in $G$.

At this point, all the neighbors of $S'$ contribute 0 to the
\InfDiffMax objective function (because they are reachable under $E'$
already), and the objective value of $S'$ is exactly $|S'| = |S|$.
\end{extraproof}

\section{Experiments}
\label{sec:experiments}
While we saw in Section~\ref{sec:instability-example} that examples
highly susceptible (with errors of magnitude $\Omega(n)$) to small
perturbations exist, the goal of this section is to evaluate
experimentally how widespread this behavior is for realistic social
networks.

\subsection{Experimental Setting}
We carry out experiments under the \ICModel, for six classes of
graphs --- four synthetic and two real-world.
In each case, the model/data give us a simple graph or multigraph.
Multigraphs are converted to simple graphs by collapsing parallel
edges to a single edge with weight $\ExpEdgeCD{e}$ equal to the number
of parallel edges;
for simple graphs, all weights are $\ExpEdgeCD{e} = 1$.
The observed probabilities for edges are
$\ActProbD{e} = \ExpEdgeCD{e} \cdot \Scaling$;
across experiments, we vary the base probability $\Scaling$ to take
on the values $\Set{0.01,0.02,0.05,0.1}$.
The resulting parameter vector is denoted by $\AllParamValuesObs$.

The uncertainty interval for $e$ is
$\ID{e}=[(1-\ExpInfDet) \ActProbD{e}, (1+\ExpInfDet) \ActProbD{e}]$;
here, $\ExpInfDet$ is an uncertainty parameter for the estimation,
which takes on the values $\Set{1\%,5\%,10\%,20\%,50\%}$ in our
experiments.
The parameter vectors $\AllParamValuesMax$ and $\AllParamValuesMin$
describe the settings in which all parameters are as large (as small,
respectively) as possible.

\subsection{Network Data}
We run experiments on four synthetic networks and two real social networks.
Synthetic networks provide a controlled environment in which to
compare observed behavior to expectations, while real social networks
may give us indications about the prevalence of vulnerability to
perturbations in real networks that have been studied in the past.

\textbf{Synthetic Networks.}
We generate synthetic networks according
to four widely used network models.
In all cases, we generate undirected networks with 400 nodes.
The network models are: (1) the 2-dimensional grid, (2) random regular graphs,
(3) the Watts-Strogatz Small-World (SW) Model \cite{watts:strogatz} on a
ring with each node connecting to the 5 closest nodes on each side
initially, and a rewiring probability of 0.1.
(4) The Barab\'{a}si-Albert Preferential Attachment (PA)
Model \cite{barabasi:albert:emergence} with 5 outgoing edges per node.
For all synthetic networks, we select $k=20$ seed nodes.

\textbf{Real Networks.}
We consider two real networks to evaluate the susceptibility of
practical networks: one (\emph{STOCFOCS}) is a co-authorship network
of theoretical CS papers; the other (\emph{Haiti}) is a Retweet network.

The co-authorship network, \emph{STOCFOCS}, is a multigraph extracted
from published papers in the conferences STOC and FOCS from 1964--2001.
Each node in the network is a researcher with at least one publication
in one of the conferences.
For each multi-author paper, we add a complete undirected graph among
the authors. As mentioned above, parallel edges are then compressed
into a single edge with corresponding weight.
The resulting graph has $1768$ nodes and $10024$ edges.
Due to its larger size, we select $50$ seed nodes.

The \emph{Haiti} network is extracted from tweets of 274 users on the
topic \emph{Haiti Earthquake} in Twitter.
For each tweet of user $u$ that was retweeted by $v$, we add a
directed edge $(u,v)$. We obtain a directed multigraph;
after contracting parallel edges, the directed graph has 383
weighted edges.
For this network, due to its smaller size, we select $20$ seeds.

In all experiments, we work with uniform edge weights
$\Scaling$, since --- apart from edge multiplicities --- we have no
evidence on the strength of connections. It is a promising direction
for future in-depth experiments to use influence strengths inferred
from real-world cascade datasets by network inference methods such as
\cite{gomez-rodriguez:balduzzi:schoelkopf:uncovering,goyal:bonchi:lakshmanan:learning,myers:leskovec:convexity}.

\subsection{Algorithms}
\label{sec:random-greedy}

Our experiments necessitate the solution of two algorithmic problems:
Finding a set of size $k$ of maximum influence, and finding a set of
size $k$ maximizing the \emph{influence difference}. The former is a
well-studied problem, with a monotone submodular objective function.
We simply use the widely known $1-1/e$ approximation algorithm due to
Nemhauser et al.~\cite{nemhauser:wolsey:fisher}, which is best
possible unless P=NP.

For the goal of \InfDiffMax, we established (in
Section~\ref{sec:ApxProof}) that the objective function is hard to
approximate better than a factor $O(n^{1-\epsilon})$ for any
$\epsilon > 0$. For experimental purposes, we use the
\emph{Random Greedy} algorithm of Buchbinder et
al.~\cite{Buchbinder:Feldman:Naor:Schwartz}, given as
Algorithm~\ref{Alg:RandomGreedy} below.
It is a natural generalization of the simple greedy algorithm of
Nemhauser et al.: Instead of picking the \emph{best} single element to
add in each iteration, it first finds the set of the $k$ individually
best single elements (i.e., the elements which when added to the
current set give the largest, second-largest, third-largest, $\ldots$,
\Kth{k}-largest gain).
Then, it picks one of these $k$ elements uniformly at random and continues.

This particular choice of algorithm was motivated by an incorrect
claim included in a prior version of this work, namely, that the
\InfDiffMax objective is (non-monotone) submodular.
For such functions, the Random Greedy algorithm guarantees at least an
0.266-approximation, and the guarantee improves to nearly $1/e$ when
$k \ll n$. Furthermore, the Random Greedy algorithm is simpler and
more efficient than other algorithms with slightly superior
approximation guarantees. We stress that these guarantees are not
obtained for our objective function, as submodularity does not hold.

\begin{algorithm}[h]
\begin{algorithmic}[1]
\STATE Initialize: $S_0\leftarrow\emptyset$
\FOR{$i=1, \ldots, k$}
\STATE Let $M_i \subseteq V \setminus S_{i-1}$ be the subset of size $k$ maximizing
$\sum_{u \in M_i} g(S_{i-1} \cup \Set{u}) - g(S_{i-1})$.
\STATE Draw $u_i$ uniformly at random from $M_i$.
\STATE Let $S_i \leftarrow S_{i-1} \cup \Set{u_i}$.
\ENDFOR
\STATE Return $S_k$
\end{algorithmic}
\caption{Random Greedy Algorithm \label{Alg:RandomGreedy}}
\end{algorithm}

The running time of the Random Greedy Algorithm is $O(kC|V|)$, where
$C$ is the time required to estimate $g(S \cup \Set{u})-g(S)$.
In our case, the objective function is \#P-hard to evaluate exactly
\cite{chen:wang:wang:prevalent,chen:yuan:zhang:scalable}, but
arbitrarily close approximations can be obtained by Monte Carlo simulation.
Since each simulation takes time $O(|V|)$, if we run
$M = 2000$ iterations of the Monte Carlo simulation in each
iteration, the overall running time of the algorithm is $O(k M |V|^2)$.

A common technique for speeding up the greedy algorithm for maximizing
a submodular function is the CELF heuristic of Leskovec et al.~\cite{LKGFVG}.
When the objective function is submodular, the standard greedy algorithm and
CELF obtain the same result. However, when it is not, the results may
be different. In the previous version of this article, we had used the
CELF heuristic due to the incorrect belief that the objective function
was submodular.
In this revised version, we instead report the results from rerunning
all the experiments without the use of the CELF heuristic.
The single exception is the largest input, the STOCFOCS network.
(Here, the greedy algorithm without CELF did not finish
in a reasonable amount of time.)
For all networks other than STOCFOCS, the results using CELF are not
significantly different from the reported results without the CELF
optimization. For STOCFOCS, we instead report the result including the
CELF heuristic.


\subsection{Results}
In all our experiments, the results for the Grid and Small-World network
are sufficiently similar that we omit the results for grids here.
As a first sanity check, we empirically computed
$\max_{S: |S| = 1} \IDMFuncD{\AllParamValuesMax,\AllParamValuesMin}{S}$
for the \emph{complete graph} on 200 nodes with
$\ID{e} = [1/200 \cdot (1-\ExpInfDet), 1/200 \cdot (1+\ExpInfDet)]$ and $k=1$.
According to the analysis in Section~\ref{sec:instability-example},
we would expect extremely high instability.
The results, shown in Table~\ref{Tab:cliqueEx}, confirm this expectation.
\begin{table}[h]
\centering
\begin{tabular}{|c||c|c|}
  \hline
  $\ExpInfDet$ & $\IMFuncSym[\AllParamValuesMax]$ & $\IMFuncSym[\AllParamValuesMin]$ \\\hline\hline
  $50\%$ & $66.529$ & $1.955$ \\\hline
  $20\%$ & $23.961$ & $4.253$ \\\hline
  $10\%$ & $15.071$ & $6.204$ \\\hline
\end{tabular}
\caption{Instability for the clique $K_{200}$.}\label{Tab:cliqueEx}
\end{table}

\begin{figure*}[t]
  \centering
  \begin{tabular}{cccc}
 \includegraphics[width=0.23\textwidth]{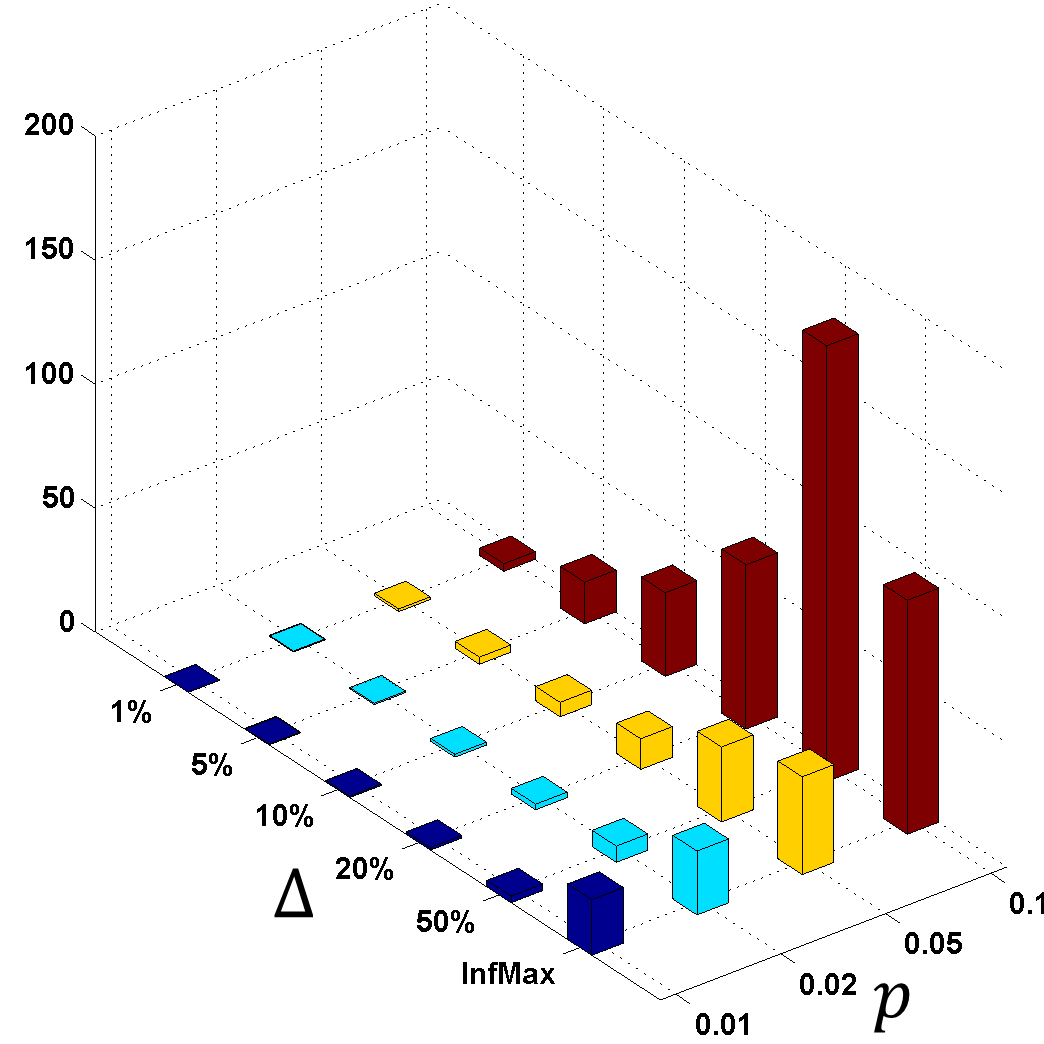}& \includegraphics[width=0.23\textwidth]{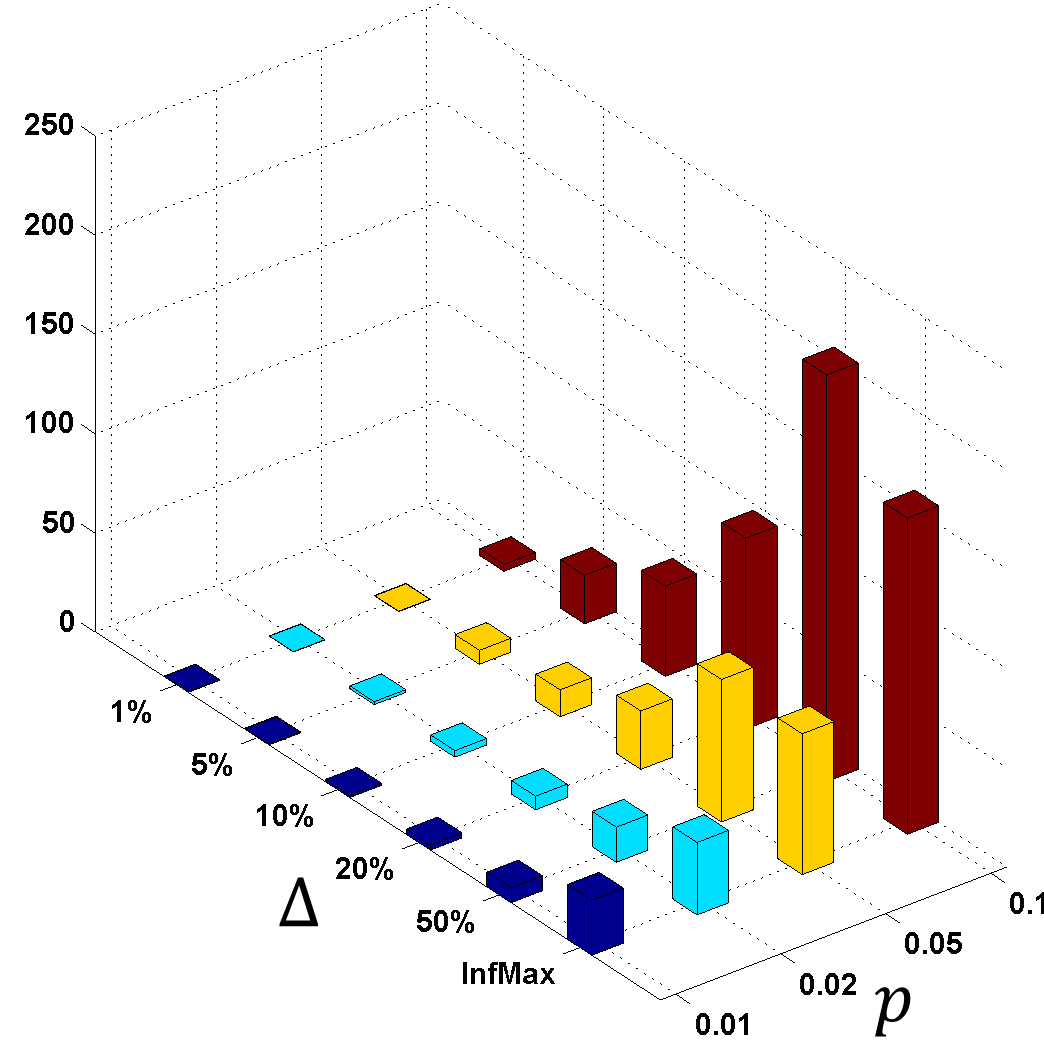}&
 \includegraphics[width=0.23\textwidth]{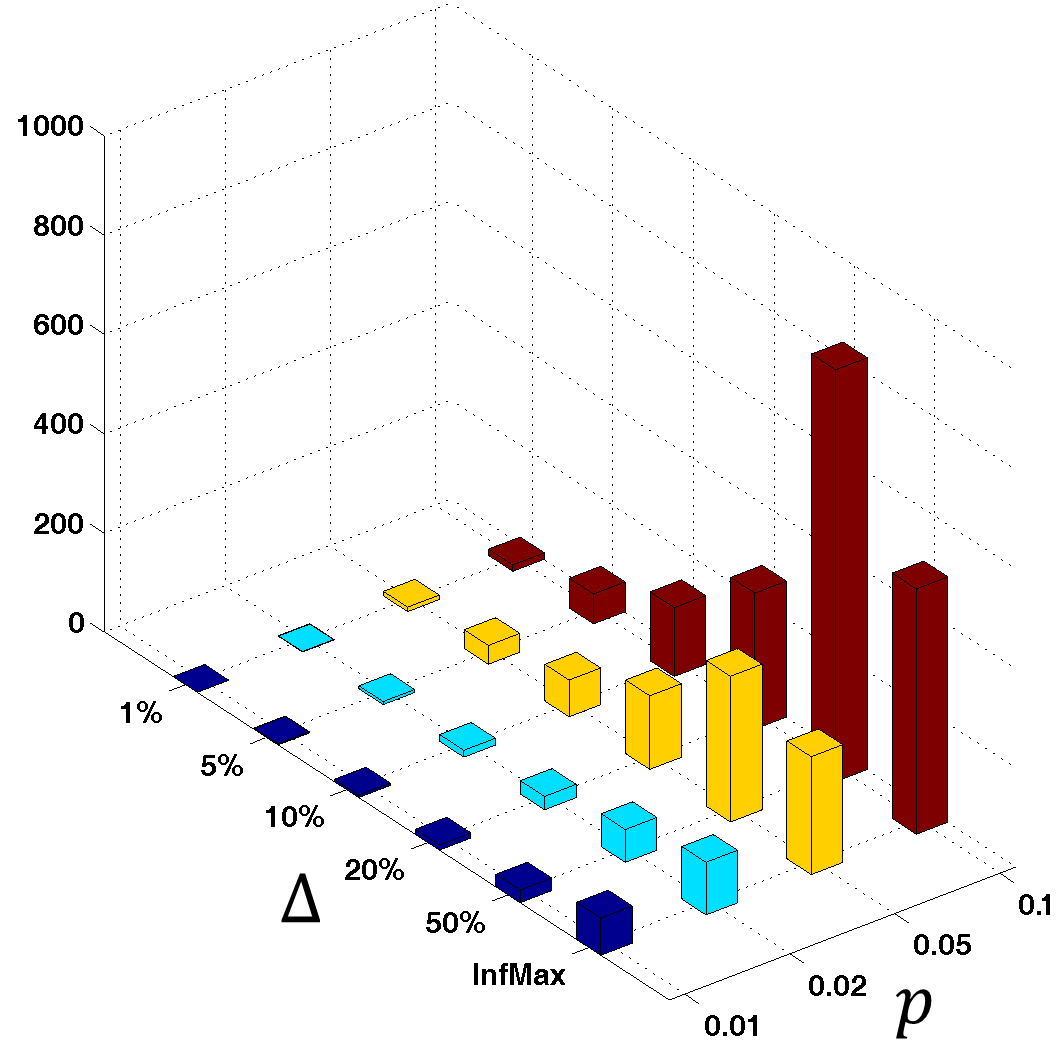}& \includegraphics[width=0.23\textwidth]{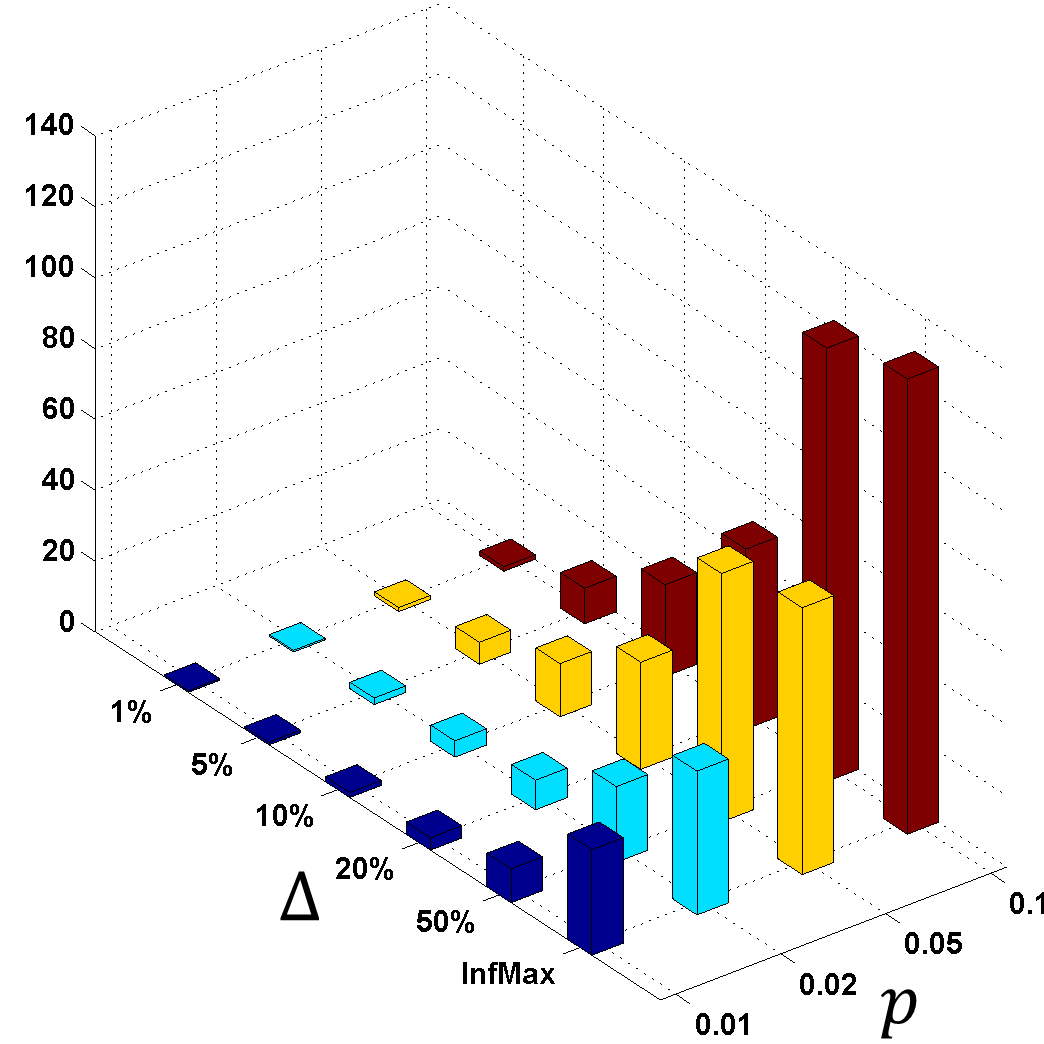} \\
       (a) Small World & (b) PA & (c) STOCFOCS & (d) Haiti\\
  \end{tabular}
  \caption{Comparison between \InfDiffMax and \InfMax results for four different networks. (The result of STOCFOCS network is obtained with CELF optimization.)}\label{Fig:CompareAll}
\end{figure*}

Next, Figure~\ref{Fig:CompareAll} shows the
(approximately) computed values
$\max_{S: |S| = k} \IDMFuncD{\AllParamValuesMax,\AllParamValuesMin}{S}$,
and --- for calibration purposes ---
$\max_{\SeedS: |\SeedS| = k} \IMFuncD{\AllParamValuesObs}{\SeedS}$
for all networks and parameter settings.
Notice that the result is obtained by running the Random Greedy
algorithm without any approximation guarantee.
However, as the algorithm's output provides a lower bound on the
maximum influence difference, a large value suggests that \InfMax
could be unstable. On  the other hand, small values \emph{do not}
guarantee that the instance is stable, as the algorithm provides no
approximation guarantee.

While individual networks vary somewhat in their susceptibility, the
overall trend is that larger estimates of baseline probabilities
$\ActProb$ make the instance more susceptible to noise, as do
(obviously) larger uncertainty parameters $\ExpInfDet$.
In particular, for $\ExpInfDet \geq 20\%$, the noise (after scaling)
dominates the \InfMax objective function value, meaning that
optimization results should be used with care.

\begin{figure*}
  \centering
  \begin{tabular}{ccc}
    \includegraphics[width=0.31\textwidth]{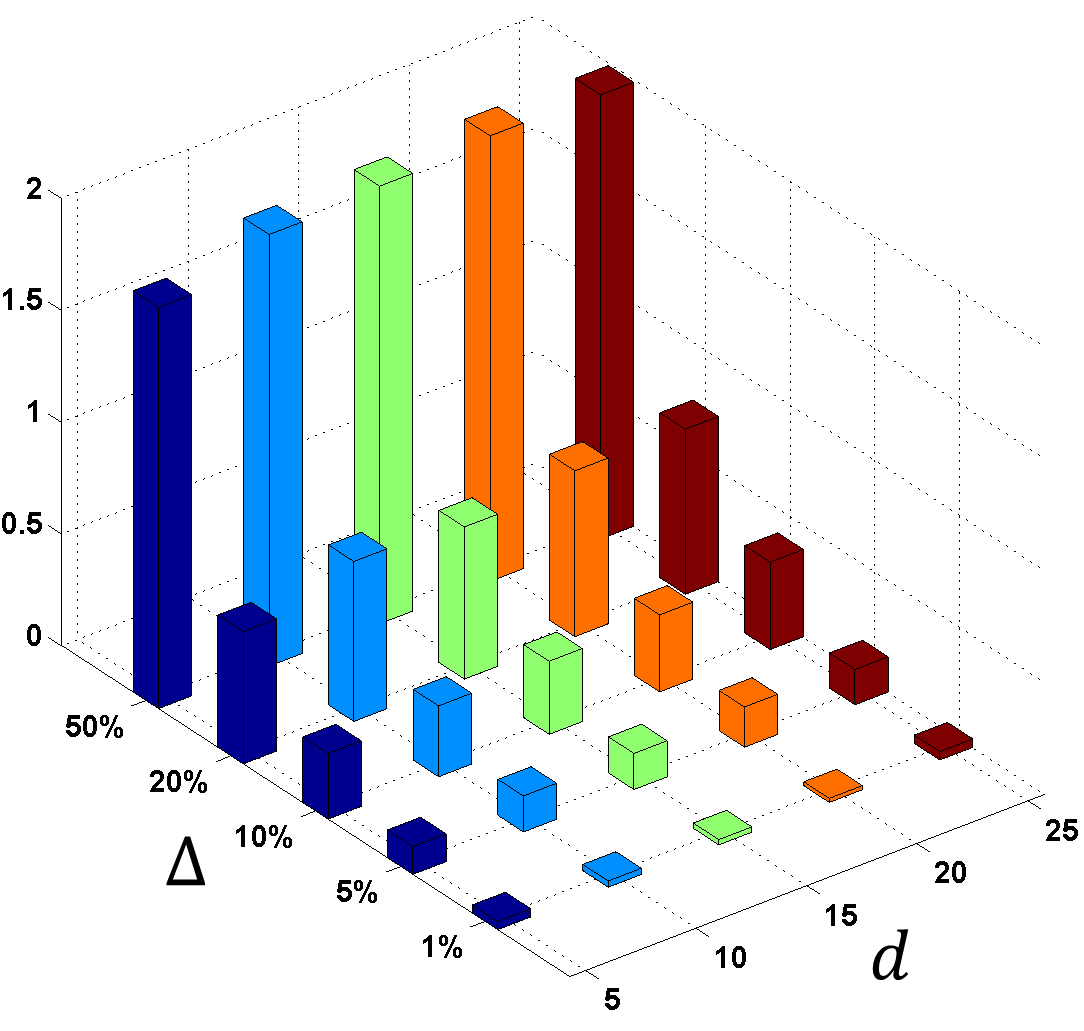} &
    \includegraphics[width=0.31\textwidth]{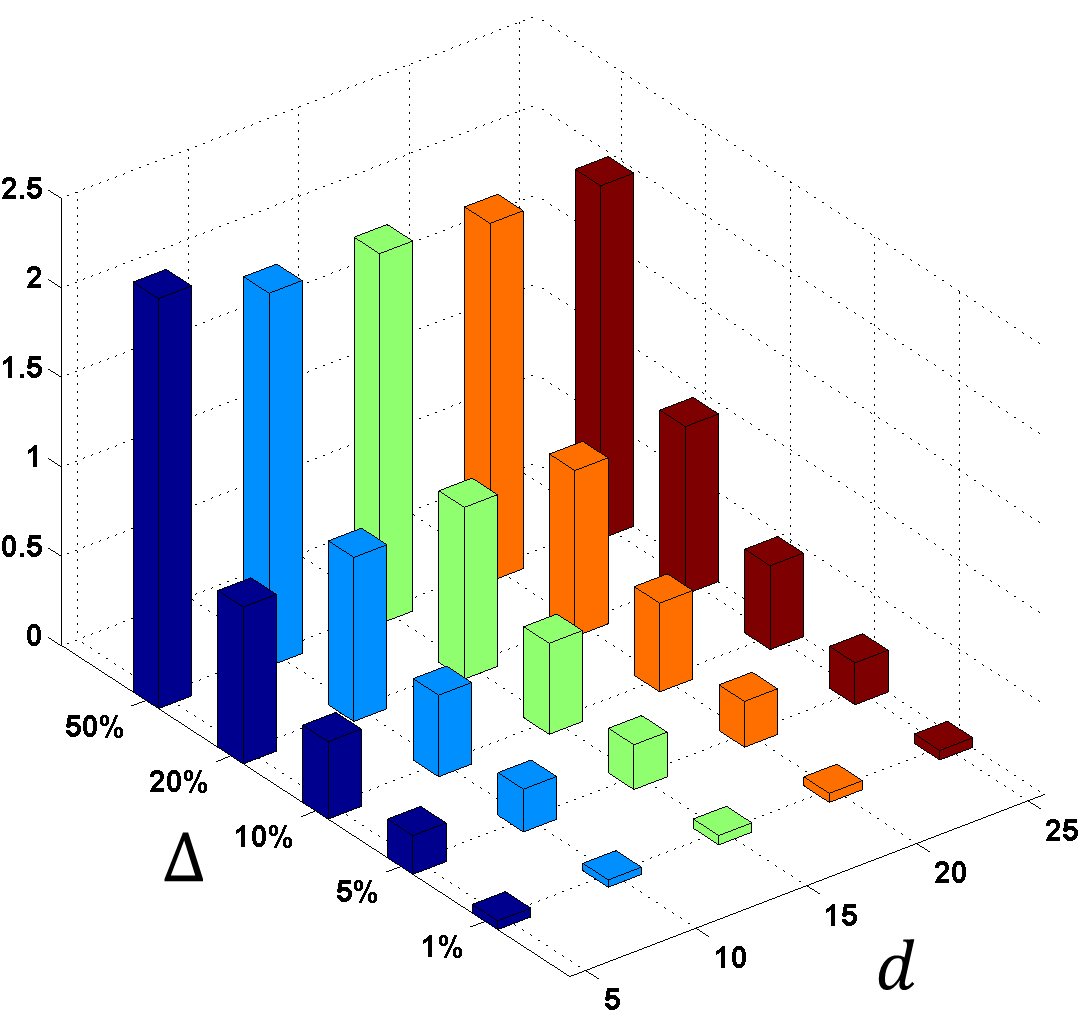} &
    \includegraphics[width=0.31\textwidth]{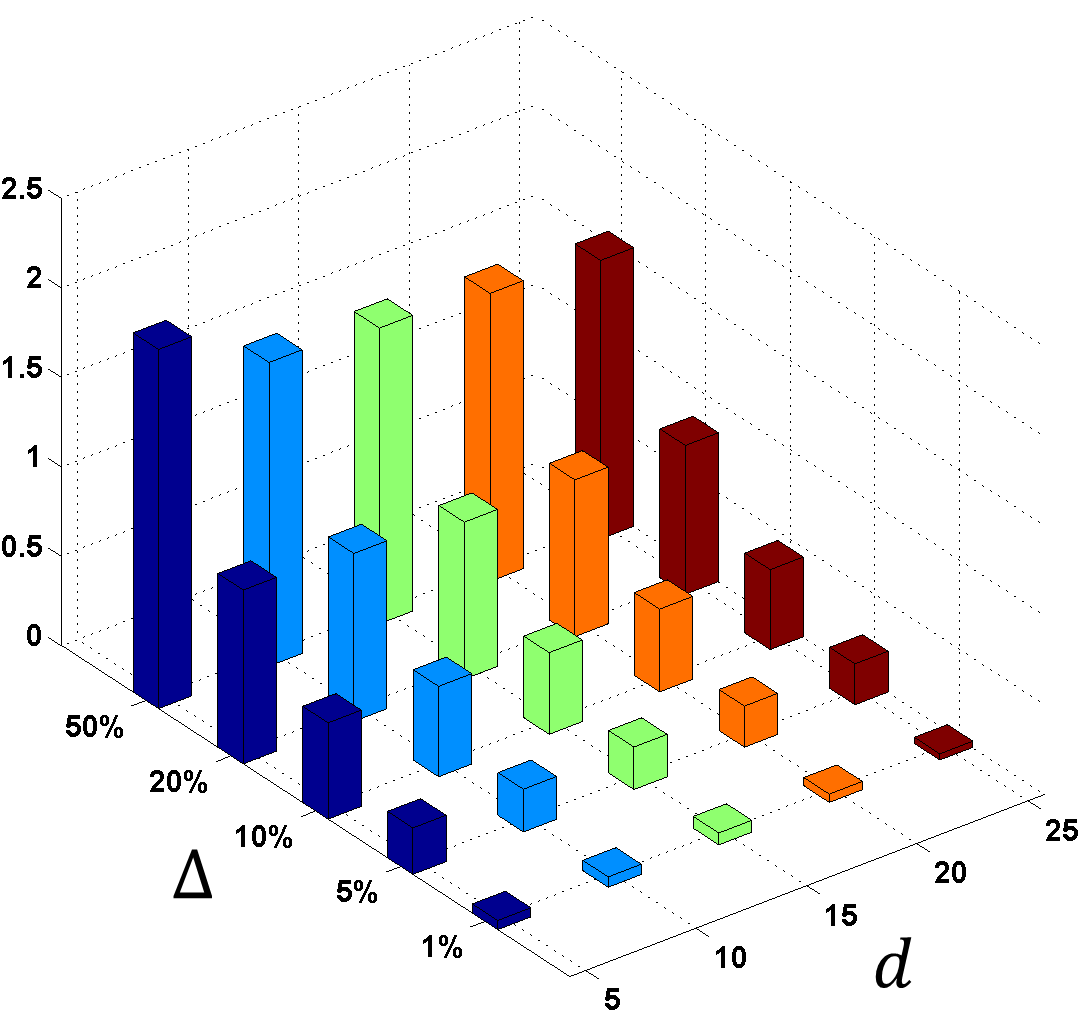} \\
    $\alpha=-20\%$ & $\alpha=0$ & $\alpha=20\%$ \\
  \end{tabular}
  \caption{Ratio between the computed values of \InfDiffMax and \InfMax under random regular graphs with different degree.}\label{Fig:regCompare}
\end{figure*}

Next, we evaluate the dependence of the noise tolerance on the
degrees of the graph, by experimenting with random $d$-regular graphs
whose degrees vary from $5$ to $25$. It is known that such graphs are
expanders with high probability, and hence have
percolation thresholds of $1/d$ \cite{alon:benjamini:stacey}.
Accordingly, we set the base probability to $(1+\alpha)/d$ with
$\alpha\in\{-20\%,0, 20\%\}$.
We use the same setting for uncertainty intervals as in the previous
experiments.
Figure~\ref{Fig:regCompare} shows the ratio between \InfDiffMax and
\InfMax, i.e.,
$\frac{\max_{S} \IDMFuncD{\AllParamValuesMax,\AllParamValuesMin}{S}}{\max_{S} \IMFuncSym[\AllParamValuesObs](S)}$, with $\alpha\in\{-20\%,0, 20\%\}$.
It indicates that for random regular graphs,
the degree does not appear to significantly affect stability, and that
again, noise around 20\% begins to pose a significant challenge.
Moreover, we observe that the ratio reaches its
minimum when the edge activation probability is exactly at the
percolation threshold $1/d$.
This result is in line with percolation theory and also the analysis
of Adiga et al.~\cite{adiga:kuhlman:sensitivity}.

As a general takeaway message, for larger amounts of noise (even just
a relative error of 20\%) --- which may well occur in practice --- a
lot of caution is advised in using the results of algorithmic \InfMax.

\section{Discussion}
\label{sec:conclusions}

We began a study of the stability of \InfMax when the
input data are adversarially noisy.
We showed that estimating the susceptibility of an instance to
perturbations can be cast as an \InfDiffMax problem.
Unfortunately, the \InfDiffMax problem under the \ICModel is
as hard to approximate as the \textsc{Independent Set}
problem.
While we do not at present have a comparable approximation hardness
result for the \LTModel, we consider it unlikely that the \InfDiffMax
objective could be much better approximated for that model.

We used the Random Greedy algorithm of Buchbinder et al.~to gain an
empirical understanding of the prevalence of instability
on several synthetic and real networks.
The results suggest that 20\% relative error could lead to a significant
risk of suboptimal outputs. Given the noise inherent in all estimates
of social network data, this suggests applying extreme caution before
relying heavily on results of algorithmic \InfMax.

The fact that our main theorem is negative (i.e., a strong
approximation hardness result) is somewhat disappointing, in that it
rules out reliably categorizing data sets as stable or unstable.
This suggests searching for models which remain algorithmically
tractable while capturing some notion of adversarially perturbed
inputs. The issue of noise in social network data will not
disappear, and it is necessary to understand its impact more
fundamentally.

While we begin an investigation of how pervasive susceptibility to
perturbations is in \InfMax data sets, our investigation is
necessarily limited. Ground truth data are by
definition impossible to obtain, and even good and reliable inferred
data sets of actual influence probabilities are currently not
available. The values we assigned for our experimental evaluation
cover a wide range of parameter values studied in past work, but the
community does not appear to have answered the question whether these
ranges actually correspond to reality.

At an even more fundamental level, the \emph{models} themselves have
received surprisingly little thorough experimental validation, despite
having served as models of choice for hundreds of papers over the last
decade. In addition to verifying the susceptibility of models to
parameter perturbations, it is thus a pressing task to verify how
susceptible the optimization problems are to incorrect models.
The verification or falsification of sociological models for
collective behavior likely falls outside the expertise of the computer
science community, but nonetheless needs to be undertaken before any
significant impact of work on \InfMax can be truthfully claimed.

\subsubsection*{Acknowledgments}
We are deeply indebted to Debmalya Mandal, Jean
Pouget-Abadie and Yaron Singer for bringing to our attention a
counter-example to a theorem incorrectly claimed in the previous
version of this article.

Furthermore, we would like to thank Shaddin Dughmi for useful pointers and
feedback, Shishir Bharathi and Mahyar Salek for useful
discussions, and anonymous reviewers for useful feedback on prior versions.
Xinran He was supported in part by the grant ONR MURI W911NF-11-1-0332
and DARPA SMISC W911NF-12-1-0034.

\bibliographystyle{abbrv}

\end{document}